\documentclass{iopjournal}
\usepackage{draftwatermark}
\SetWatermarkText{PREPRINT} % The text of your watermark
\SetWatermarkScale{1.2}     % Adjusts the size (1.0 is default)
\SetWatermarkColor[gray]{0.85} % A light gray that won't make text hard to read
\usepackage[backend=biber, style=phys]{biblatex}  %backend=bibtex, or backend=biber is 'better'  
\expandafter\let\csname equation*\endcsname\relax
\expandafter\let\csname endequation*\endcsname\relax
\usepackage{amsmath, amsbsy}

\usepackage{lipsum,multicol} % for wide equations spanning 2 columns

\usepackage [english]{babel}
\usepackage [autostyle, english = american]{csquotes}
\MakeOuterQuote{"}

\usepackage{subcaption}

\usepackage{amssymb}

\usepackage{graphicx}

\usepackage[final]{changes}

\begin{document}

\articletype{Paper} %	 e.g. Paper, Letter, Topical Review...

\title{Tearing Stability Prediction Combining Toroidal Calculations With a Two-Fluid Slab Layer}

\author{D.A. Burgess$^1$\orcid{0000-0002-5997-8324}, N.C. Logan$^1$\orcid{0000-0002-3268-7359}, 
J.-K. Park$^{2}$\orcid{0000-0003-2419-8667},
C. Paz-Soldan$^{1}$\orcid{0000-0001-5069-4934},
}

\affil{$^1$Columbia University, New York, New York 10027, USA}

\affil{$^2$Seoul National University, Seoul, Korea}

\email{daniel.burgess@columbia.edu}

\begin{abstract}

A new classical TM stability simulation workflow has been developed that solves the resistive inner-layer equations in a plasma slab to yield a linear, quasi-toroidal TM growth rate $\gamma$ and mode rotation frequency $\omega$. This workflow combines two-fluid and drift MHD effects in a slab approximation of the resistive inner layer (SLAYER) with an effective tearing stability index as $\Delta(\gamma,\omega) = \Delta' - \Delta_\mathrm{crit}$. SLAYER is used to calculate the inner-layer $\Delta(\gamma,\omega)$, the STRIDE code is used to calculate a toroidal $\Delta'$ that includes shaping effects, and the toroidal $\Delta_\mathrm{crit}$ incorporates effects of thermal conduction on Glasser stabilization. This workflow is rapid and numerically robust across reactor-relevant plasma conditions, and yields growth rates that closely align with analytic predictions in well-documented linear growth rate regimes. Using synthetic equilibria, TM stability was calculated across scans of plasma $\beta$, inverse aspect-ratio, and toroidal current profile gradient. These scans effectively benchmarked this STRIDE+SLAYER workflow against existing models and showed reliable stability predictions in shaped H-mode-like plasmas. This capability to quickly and robustly predict classical tearing stability in tokamaks will facilitate the mapping of TM stable operational regimes and design of safe discharge trajectories in future devices.

\end{abstract}

%\begin{multicols}{2} % <--- Start the two-column layout here
%\noindent{\it LLNL-JRNL-821772}
% \submitto{\NF}
%\maketitle
% 

%\tableofcontents{}


\section{\label{sec:Motivation} Motivation}

Tearing modes (TMs) are a class of resistive magnetohydrodynamic (MHD) instability that occur in toroidal magnetic confinement fusion configurations with a significant plasma current, such as tokamaks. While TMs can be benign, they can also pose a serious threat to stable plasma operation -- as the TM opens a toroidally periodic chain of magnetic islands and their widths grow, particles and thermal energy will be lost from the inboard to the outboard side of the islands, causing deleterious transport \cite{LaHaye_2006_PoP, sweeney_statistical_2017}. In severe cases, larger islands can ``lock'' to the wall due to eddy currents, drag against the plasma's toroidal rotation, and cause a disruption \cite{deVries_2011-NF}. Disruptions involve a complete loss of the plasma, can damage the tokamak, and must be minimized in a fusion power plant \cite{sweeney2020mhd, lehnen2015disruptions}. For these reasons, understanding and if possible, predicting the onset of TMs will help design safer discharges from ramp-up to flat-top and ramp-down, substantially aiding stable tokamak operation. 

TMs grow on surfaces of closed field lines in the plasma, known as ``resonant surfaces'' and defined by a rational value of the safety factor $q=m/n$, where $m$ is the number of poloidal field line transits and $n$ the number of toroidal transits. The global ideal MHD force balance solution is singular at rational $q=m/n$, and the size of this singular jump is related to the strength of a driving force exerted by the toroidal current density gradient across this resonant surface in the radial direction. This force is associated with the free energy available for tearing this rational surface into a chain of magnetic islands, and can be measured by the ideal MHD tearing stability parameter $\Delta'(m,n)$ \cite{furth_tearing_1973}. However, resistivity is required to calculate the linear growth rates and rotation frequencies of modes on rational surfaces, as well as to model TM stabilization by curvature effects (Ref.~\cite{GGJ_finiteP}; see Section~\ref{sec:STRIDE_intro}), therefore $\Delta'$ lacks certain physics for more predictive use. Several ``straight-through'' code suites have been developed which treat the entire plasma volume using resistive MHD, such as NIMROD \cite{glasser1999nimrod} or M3D-C1 \cite{Ferraro_2010-PoP}. However, a more computationally efficient approach is to treat only a narrow region on either side of the rational surface with more detailed resistive or two-fluid physics, assume ideal MHD everywhere else in the plasma, and match the ideal and resistive solutions to solve for the mode growth rate \cite{fitzpatrick_2023}. This asymptotic matching approach is well known and has been successfully demonstrated in e.g. the RDCON code, which utilizes a simple inner region model and retains surface-to-surface coupling \cite{Glasser_RDCON_2016}. 

The novel approach taken in this work uses a newer, well-established and detailed slab model of the resistive inner region \cite{Fitzpatrick_Waelbroeck_2005, Cole_Fitzpatrick_2006} (implemented numerically in the SLAYER code \cite{JKP_SLAYER_2022}) and solves the canonical dispersion relation for TM growth by asymptotically matching this inner region solution with a fully toroidal ideal MHD $\Delta'$ from the STRIDE code \cite{Glasser_STRIDE_2018}. Mode stabilization due to curvature and thermal transport effects is additionally included, which may be an important modification for model relevance in present day and future tokamaks. The STRIDE+SLAYER TM growth rates align with analytic theory and the TJ code across a wide parameter space, and differing stability predictions at low aspect ratio are observed (a regime where STRIDE can be expected to be more accurate). Growth rate calculations are additionally made in a shaped H-mode-like plasma with multiple rational surfaces and self-consistently calculated bootstrap current, showing applicability to experimental discharges in existing and future devices. 

In Section~\ref{sec:SLAYER_workflow}, a procedure is outlined for calculating the inner region solutions, then the outer region solutions, and finally a method for combining the two into a dispersion relation and solving for the TM growth rate. In Section~\ref{sec:analytic_benchmark}, 
STRIDE+SLAYER growth rates are compared against analytic growth rates across four established linear growth rate regimes. In Section~\ref{sec:TJ_benchmark}, this STRIDE+SLAYER workflow is benchmarked against the TJ code on both a plasma $\beta$ scan and an inverse aspect-ratio scan. In Section~\ref{sec:D3D_example}, the STRIDE+SLAYER workflow is applied to a shaped, synthetic equilibrium with realistic H-mode-like kinetic and toroidal current profiles. Finally, in Section~\ref{sec:summary}, findings and future work are discussed.

%Introduce tearing modes. (Cite De Vries). Introduce motivation for matching ideal outer and resistive inner regions (full nonlinear is hard). Explain why $\Delta'$ may be insufficient for tearing stability predictions and summarize importance of $\Delta_{\mathrm{eff}}$ for stability and growth rates, mode rotations. Extremes exist in simple perturbative analysis and full non linear, we want to span the two and still incorporate important effects like shaping. Positioning this code within the current suite of options. More citations: Turco '24, Boyes '24, Kim '25. Introduce SLAYER. Introduce inner-outer layer matching. Brief section-by-section paper overview.

\section{\label{sec:SLAYER_workflow} SLAYER Tearing Stability Workflow}

\subsection{\label{sec:SLAYER_intro} Calculation of the Inner-Layer $\Delta$}

The tearing of a magnetic surface into a magnetic island is a fundamentally resistive process, and many physical effects can influence whether a mode amplitude at a rational surface will be damped or grow exponentially into a nonlinear island of finite width. A variety of analytic inner-layer models of varying fidelity have been developed which incorporate relevant physics effects at a rational surface, and both resistive and inertial effects have been successfully addressed even in a complete toroidal geometry. However, higher order two-fluid MHD effects such as drifts, the presence of finite rotation, long mean-free-path physics, semi-collisional effects, and Hall effects have thus far only been implemented in a simplified 3D plasma slab geometry \cite{JKP_SLAYER_2022}. A slab model is an appropriate approximation of the narrow resistive region around a rational surface as long as the region remains narrow (i.e., high Lundquist number $S$). Additionally, the majority of relevant geometric effects are captured in the outer region solution, which is computed here in a full toroidal geometry.

There are certain caveats to modeling the inner region with a slab model, particularly a reduced slab model that is analytically tractable. These include the assumption of no pressure gradient and no parallel flow in the inner region. The neglect of pressure gradients is justifiable in the case of a sufficiently narrow resistive layer width, which is a reasonable approximation in $\sim$keV tokamaks (the layer width scales roughly as $T_e^{-3/5}$ \cite{Furth_1963_PoF}). The neglect of parallel flow is acceptable given that the plasma is not high $\beta \propto P/B_t^2$ ($c_\beta \gtrsim 0.2$) \cite{Lee_2024_NF}, which is a reasonable assumption in present-day and future tokamaks as high $\beta$ operation increases neoclassical tearing mode (NTM) risk \cite{Buttery_2000_PPCF}. 

The inner-layer response to a mode amplitude can be investigated through a scalar quantity termed the inner-layer $\Delta$ \cite{fitzpatrick_2023}, which is obtained by solving the two-fluid drift MHD equations in a slab geometry. The slab model used in SLAYER is the well-known analytic four field model developed by Fitzpatrick, Cole, and Waelbroeck \cite{Fitzpatrick_Waelbroeck_2005, Cole_Fitzpatrick_2006}. The four field model equations are reduced via variable substitution, a generalized Fourier-Laplace transform, and a Riccati transform into a single analytic, rapidly solvable ODE (for a full derivation of this ODE, see Refs. \cite{Cole_Fitzpatrick_2006} and \cite{JKP_SLAYER_2022}). The full set of SLAYER ODE inputs for a given rational surface $(Q$, $Q_e$, $Q_i$,$C$,$D$,$P$,$\tau)$ are defined in Table~\ref{tab:normalized_variables}. The real and imaginary components of $Q$ correspond to the rotation and growth rate of the mode amplitude, respectively.

\begin{table}[t]
\centering
\caption{SLAYER normalized input variables (top), sub-variables (bottom), and their definitions. For $C^2$ and $P$, on the left are the original Fitzpatrick, Cole, \& Waelbroeck definitions retained in the original SLAYER release \cite{JKP_SLAYER_2022}, on the right are equivalencies used in other recent implementations \cite{fitzpatrick_2022, fitzpatrick_TJ_2025}. $\omega_{\ast\,e}$ and $\omega_{\ast\,i}$ are the electron and ion diamagnetic frequencies, respectively. For a complete list of variables defined from plasma parameters, see Appendix~\ref{appen:plasma_params}.}
\begin{tabular}{l|l|p{8cm}}
\multicolumn{1}{c}{\textbf{Variable}} & \multicolumn{1}{c}{\textbf{Definition}} & \multicolumn{1}{c}{\textbf{Description}} \\
\hline\hline
$\mathrm{Re}(Q)$ & $ \hat \omega = S^{1/3} \omega \tau_H$ & Normalized mode rotation frequency (in local E$\times$B frame)\\
\hline
$Im(Q)$ & $\hat \gamma = S^{1/3} \gamma \tau_H$ & Normalized growth rate (in local E$\times$B frame)\\
\hline
$Q_e$ & $S^{1/3} \omega_{\ast\,e} \tau_H$ & Normalized electron diamagnetic frequency \\
\hline
$Q_i$ & $S^{1/3} \omega_{\ast\,i} \tau_H$ & Normalized ion diamagnetic frequency \\
\hline
$D$ & $S^{1/3}\,\left[ \omega_{\ast\,e}/(\omega_{\ast\,e}-\omega_{\ast\,i}) \right]^{1/2} ({d}_\beta/r_s)$ & Normalized ion skin depth \\
\hline
$C^2$ & $(c_{\beta}^2 + (1-c_{\beta}^2)K) \simeq P_\perp = (\tau_R \chi_\perp)/r_s^2$ & Resistive diffusion $\simeq$ perp. magnetic Prandtl number \\
\hline
$P$ & $\tau_R/\tau_V \simeq P_\phi = (\tau_R \chi_\phi)/r_s^2$ & Norm. ion viscosity $\simeq$ toroidal magnetic Prandtl number \\
\hline
$\tau$ & $T_i/T_e$ & Ion to electron temperature ratio \\
\end{tabular}
\label{tab:normalized_variables}
\end{table}

Using these normalized variables, the Riccati-transformed, Fourier-space ODE solved by SLAYER can be written as:

%\begin{equation}
\begin{multline}
    \frac{dW}{dp} = \left[ \frac{2p}{iq_e+p^2} - \frac{1}{p} \right]W - \frac{W^2}{p} + p\left[iq_e + p^2 \right] G(p) ,
\label{eq:dWdp}
\end{multline}
%\end{equation}

\noindent with

\begin{equation}
\begin{aligned}
    G(p) &= \dfrac{-Qq_i+iq_i(P+C^2)p^2+C^2Pp^4}{iq_e+(C^2+iq_iD^2)p^2 + (1+\tau)PD^2p^4} \\
    q_e &= Q-Q_e\\
    q_i &= Q-Q_i.
\label{eq:Gp}
\end{aligned}
\end{equation}

\noindent Here, $p$ is the Fourier frequency variable and $W$ is defined in Ref.~\cite{JKP_SLAYER_2022}. Specifying

\begin{equation}
\begin{aligned}
    a &= -(Q+iQ_e) \\
    b &= P \\
    c &= -i(Q_e-Q_i) (P / C^2) + Q + iQ_i,
\label{eq:prob}
\end{aligned}
\end{equation}

\noindent the large $p$ boundary conditions take the form 

\begin{equation}
\begin{aligned}
    W(p) = \frac{ab-c}{2\sqrt{b}}-\sqrt{P}\,p^2,
\label{eq:largeD}
\end{aligned}
\end{equation}

\noindent at $D > \frac{|C|}{P^{1/3}}$, and

\begin{equation}
\begin{aligned}
    W(p) = -1 + \left( \frac{ab-c}{2\sqrt{b}} \right)p-\sqrt{P}\,p^3
\label{eq:smallD}
\end{aligned}
\end{equation}

\noindent at $D < \frac{|C|}{P^{1/3}}$ \cite{fitzpatrick_TJ_2025}. The normalized inner-layer $\hat \Delta$ can then be obtained via a power series expansion by launching either (\ref{eq:largeD}) or (\ref{eq:smallD}) at large $p$, backwards integrating $\frac{dW}{dp}$ to small $p$, and taking

\begin{equation}
\begin{aligned}
    \lim_{p\to+0} \frac{dW}{dp} \sim \frac{\pi}{\hat \Delta} .
\label{eq:deltahat}
\end{aligned}
\end{equation}

Numerically, this limit is evaluated in SLAYER when $p$ reaches $1 \times 10^{-6}$ in the backwards integration of $\frac{dW}{dp}$.

SLAYER and the Fitzpatrick model presented in Ref. \cite{fitzpatrick_TJ_2025} differ in their definitions of anomalous diffusion, but equations \eqref{eq:dWdp}, \eqref{eq:Gp}, and \eqref{eq:deltahat} that are solved for the inner-layer $\hat \Delta$ are the same in terms of normalized variables. Comparing $C^2$ and $P$ in SLAYER with $P_{\perp}$ and $P_{\phi}$ in Ref.~\cite{fitzpatrick_TJ_2025}, it can be seen that $C^2 = P_{\perp}$ and $P = P_{\phi}$. These equivalencies are noted in Table~\ref{tab:normalized_variables}.

%This confirms that the ODEs are identical, and SLAYER is simply replacing the anomalous perpendicular energy diffusivity used by Fitzpatrick with thermal conductivity $\kappa$, and the anomalous perpendicular ion momentum diffusivity with the ion viscosity $\mu_i$. Fitzpatrick additionally sets $P_{\phi} = \tau_R / \tau_\phi$, which differs from the SLAYER definition of $P = \tau_R \eta/\chi$***.

\subsection{\label{sec:STRIDE_intro} Calculation of Outer Region $\Delta'$ and $\Delta_{\mathrm{eff}}$}

In order to obtain tearing mode growth rates, the two-fluid physics in the resistive inner region must be combined with toroidal, ideal MHD information in the outer region. In the inner-outer layer matching approach, this outer region solution is the classical tearing stability index $\Delta'$, defined as the jump in the logarithmic derivative of the radial magnetic field perturbation across a rational surface. This can be defined in a slab or cylindrical geometry as

\begin{equation}
    \Delta' = \left[\frac{1}{\psi}\frac{d\psi}{dx}\right]_{{x}=x_{s}^-}^{{x}=x_{s}^+},
\end{equation}

\noindent where $x$ is a Cartesian (in a slab geometry) or radial (cylindrical geometry) coordinate and $x_s$ is the location of a rational surface \cite{Furth_1963_PoF,Connor_2014_PPCF}. The toroidal derivation and definition of $\Delta'$ are more involved \cite{GGJ_finiteP}. The tearing stability index has been previously used as a solitary predictor of classical tearing stability with some success \cite{classical_tearing_Chu,Brennan_2002_PoP}, but it is lacking the necessary inner region resistive physics for analysis of growth rates, mode rotation frequencies, and surface-to-surface coupling. 

%$\Delta'$ is additionally a primary driver of instability in the Modified Rutherford Equation (MRE) a nonlinear equation that governs neoclassical tearing mode evolution. 

The toroidal $\Delta'$ matrix for a given equilibrium is calculated using the STRIDE code \cite{Glasser_STRIDE_2018}. $\Delta'$ is an $m \times m$ subset of the full ideal MHD $2m \times 2m$ $\mathbf{D}$ matrix, where $m$ is the number of rational surfaces analyzed in the plasma \cite{Pletzer_1994_JCP}. STRIDE includes poloidal mode coupling and fully generalized toroidal and poloidal shaping. The diagonal of the $\Delta'$ matrix provides stability indices for each rational surface, while the off-diagonal terms represent surface-to-surface coupling. In this analysis, the STRIDE+SLAYER workflow investigates the simplified, decoupled problem (described in the next section). Therefore, only the self-coupled diagonal $\Delta'$ values are used, and the off-diagonal terms are discarded.

The tearing stability index $\Delta'$ provides one measure of the free energy available in the gradient of the toroidal current profile $j_\phi$ for tearing at a given rational surface $q=m/n$. However, other physical effects beyond the $j_\phi$ gradient can further alter the available free energy for tearing. One such effect is known as Glasser stabilization (or ``GGJ'' stabilization, for Glasser-Greene-Johnson) which describes how in a toroidal geometry with good average curvature (i.e. $D_R < 0$, where $D_R$ is the resistive interchange parameter), finite pressure, and high Lundquist number S, there can be a strong stabilizing effect against tearing \cite{GGJ_finiteP}. This stabilization results from finite plasma pressure resisting the tearing of a favorably curved rational surface via sound wave propagation, and can be formulated as a critical threshold $\Delta_\mathrm{crit}$. Therefore, if $\Delta' > \Delta_\mathrm{crit}$, curvature effects will no longer stabilize against tearing. $\Delta_\mathrm{crit}$ has more recently been extended to include thermal transport effects in the vicinity of the rational surface, which can have a significant effect on the strength of curvature stabilization. A large aspect-ratio approximated $\Delta_\mathrm{crit}$ that included thermal transport effects was first derived by Ref.~\cite{curvature_Lutjens_2001}, and was later extended into a toroidal geometry by Ref.~\cite{curvature_Connor_2015}. This toroidal $\Delta_\mathrm{crit}$ was defined using the volume contained within a flux surface $V$ as the radial coordinate, but it can also be written in terms of the STRIDE and SLAYER radial coordinate $\psi_N$ (normalized poloidal flux). It is defined as

\begin{equation}
    \Delta_\mathrm{crit} = \frac{\pi^{3/2}}{2} \left( \frac{\chi_{\parallel}}{\chi_{\perp}} \right)^{1/4} V_s \left( \frac{\alpha^2 \Lambda^2}{\langle B^2 \rangle \langle  \left | \nabla V \right|^2 \rangle} \right)^{1/4} (-D_R),
\end{equation}

\noindent where $V_s$ is the volume of the singular surface, and $\langle A \rangle$ is defined as the flux average over the plasma volume \cite{curvature_Connor_2015}. 

%See~\ref{appen:toroidal_dc_coords} for the full radial coordinate transformation.

As a result, the ideal MHD tearing stability threshold $\Delta' > 0$ now becomes $\Delta_{\mathrm{eff}} = \Delta' - \Delta_\mathrm{crit} > 0$, incorporating curvature and thermal transport effects alongside the fundamental physics of toroidal current ($j_{\phi}$) profile stability. 

\subsection{\label{sec:dispersion_intro} Dispersion Relation and Calculation of Tearing Mode Growth Rates}

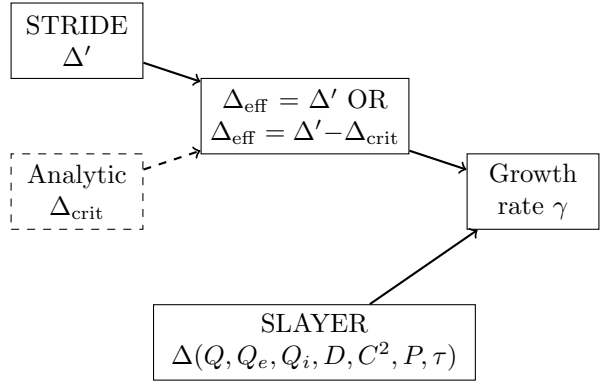
\begin{figure}[h]
\centering
\begin{tikzpicture}[
    box/.style={rectangle, draw, fill=white, text width=1.5cm, text centered, minimum height=1cm},
    optbox/.style={rectangle, draw, dashed, fill=white, text width=1.5cm, text centered, minimum height=1cm},
    arrow/.style={->, thick},
    optarrow/.style={->, thick, dashed}
]

% Left column boxes
\node[box] (delta_prime) at (-1,2) {STRIDE $\Delta'$};
\node[optbox] (delta_crit) at (-1,0) {Analytic $\Delta_\mathrm{crit}$};
\node[box, text width=4cm, anchor=west] (delta_QDP) at (0,-2) {SLAYER $\Delta(Q,Q_e,Q_i,D,C^2,P,\tau)$};

% Middle box
\node[box, text width=2.5cm] (delta_eff) at (2.0,1) {$\Delta_{\mathrm{eff}} = \Delta'$ OR $\Delta_{\mathrm{eff}} = \Delta' -\Delta_\mathrm{crit}$};

% Right box
\node[box] (omega_gamma) at (5,0) {Growth rate $\gamma$};

% Arrows
\draw[arrow] (delta_prime) -- (delta_eff);
\draw[optarrow] (delta_crit) -- (delta_eff);
\draw[arrow] (delta_eff) -- (omega_gamma);
\draw[arrow] (delta_QDP) -- (omega_gamma);

\end{tikzpicture}
\caption{STRIDE+SLAYER workflow schematic showing that a combination of outer layer tearing drive ($\Delta'$) and inner-layer response ($\Delta$) are needed to calculate the linear tearing growth rate $\gamma$. Inclusion of a proxy for Glasser stabilization ($\Delta_{\mathrm{crit}}$) in the dispersion relation is optional.}
\label{fig:flowchart}
\end{figure}

The decoupled tearing mode dispersion relation can be written as $\Delta' = \Delta(\omega,\gamma)$, where $\omega$ is the mode rotation and $\gamma$ is the growth rate, both in the local E$\times$B frame at the surface. This is a simplified case of the matrix problem described in Ref.~\cite{Pletzer_1994_JCP}, and decoupling all rational surfaces into separate dispersion relations can be justified under the assumption that significant rotation shear is present between surfaces \cite{fitzpatrick_2023}. Incorporating the $\Delta_\mathrm{crit}$ proxy for Glasser stabilization yields an effective tearing stability index $\Delta_\mathrm{eff} = \Delta'-\Delta_\mathrm{crit}$, and taking $\Delta = S^{1/3} \hat \Delta$ gives the dispersion relation solved by SLAYER:

\begin{equation}
\begin{aligned}
    \Delta'-\Delta_\mathrm{crit} = \Delta(Q),
\label{eq:dispersion}
\end{aligned}
\end{equation}

\noindent where $Q = \hat \omega + i\hat \gamma$. The inputs and output of the dispersion relation are illustrated in Fig.~\ref{fig:flowchart}. It is necessary to note that inserting a proxy for Glasser stabilization into this dispersion relation may result in spurious solutions that predict high frequency instability for large values of $\Delta_\mathrm{crit}$ (e.g. $\Delta_\mathrm{crit} >> 10$). These appear to be purely numerical modes, and since the sign of the classical tearing mode growth rate should match the sign of $\Delta_\mathrm{eff}$ \cite{fitzpatrick_2023}, these spurious solutions are discarded in this analysis.
%may not yield valid results outside of the Viscous-Resistive regime, since Glasser stabilization is inherently an inner layer effect that governs the resistive region's response to a perturbation. 
The validity of this proxy when solving for growth rates -- particularly in the experimentally relevant DR and VR regimes (see Fig.~\ref{fig:maps}) -- remains a point of ongoing work.

In order to solve this dispersion relation, a suitable $\hat \Delta$ must be found across $\hat \omega$-$\hat \gamma$ space using one's numerical method of choice. In SLAYER, a square grid scan is used for robustness (allowed by the rapidity of each ODE solve). Once a sufficiently wide $(\hat \omega,\hat \gamma)$ space has been scanned (typically $Q\pm10$), contours of $\mathrm{Re}(\Delta_{\mathrm{eff}}) = \mathrm{Re}(\Delta)$ and $\mathrm{Im}(\Delta_{\mathrm{eff}}) = \mathrm{Im}(\Delta)$ are extracted and their intersections found. The intersection with the most positive $\hat \gamma$ value not found within a pole is the dominant uncoupled growth rate for a given rational surface. Intersections at continuous values of $\Delta(Q)$ represent actual modes, whereas intersections across a pole in $\Delta(Q)$ contain an undefined value of $\Delta$ and therefore do not satisfy the dispersion relation. These contours, their intersections, and several $\Delta(Q)$ poles can be visualized across $\hat \omega$-$\hat \gamma$ space in Fig.~\ref{fig:stabscan}. 

\begin{figure}[h]
\centering\includegraphics[trim={0.5cm 0.5cm 0.0cm 0.0cm},clip,width=0.5\linewidth]{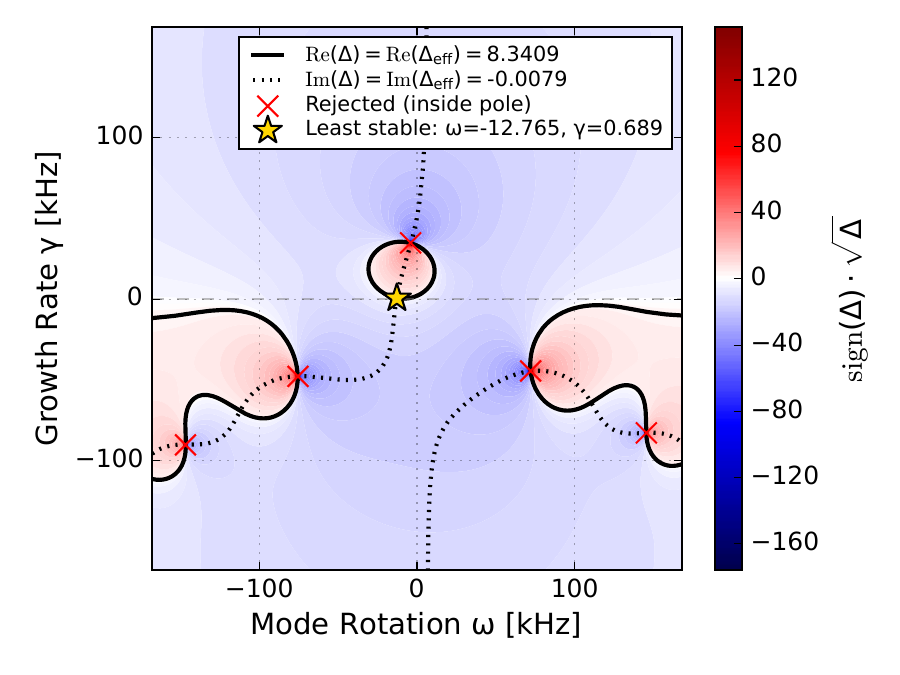}\caption{SLAYER $( \omega, \gamma)$ space growth rate root find. Shaded red and blue contours are computed using Re($\Delta$). The dominant TM growth rate (real-imaginary contour intersection) is marked with a star. Two stable solutions are also visible near $\pm110$ kHz. Poles with asymptotic $\Delta$ yield undefined solutions to the dispersion relation and are therefore rejected. \label{fig:stabscan}}
\end{figure}

In most equilibrium resonant layers tested, the solution space takes a form similar to that in Fig.~\ref{fig:stabscan}, where an unstable pole is located at $\gamma > 0$ and small $\omega$. Unstable roots only being found at small $\omega$ aligns with the reasonable expectation that a large rotation differential between the mode and plasma bulk (predominantly ExB) rotation should be stabilizing \cite{dewar1993coupled}. As $\Delta_{\mathrm{eff}}$ increases in this example (taken from the lowest $\beta$ scan point in Fig.~\ref{fig:betascan_gammas}), the contour intersection that marks the growth rate solution increases in $\gamma$ toward the pole, which aligns with the $\gamma \propto \Delta'$ scaling identified in analytic theory (see eqs. \ref{eq:DR} to \ref{eq:RI}). Conversely, as $\Delta_{\mathrm{eff}}$ decreases below $0$, the contour intersection decreases in negative $\gamma$ correspondingly. 

Further accelerating these SLAYER growth rate scans by use of an unstructured mesh of scan points and adaptive mesh refinement is planned for future work. 

\section{\label{sec:analytic_benchmark} Growth Rate Benchmark Against Analytic Linear Growth Rates}

%Show Fitzpatrick analytic linear growth rates for semi-collisional and diffusive-resistive regimes. Note that analytic linear growth rates are all stable for negative $\Delta'$. Show regime transition and full growth rate maps.

Analytic linear growth rates have previously been derived across distinct regimes that span a range of normalized total diamagnetic frequencies $Q_*$ and magnetic Prandtl numbers $P$ \cite{fitzpatrick_2022}. Using a cylindrical geometry and assuming that the plasma is surrounded by a perfectly conducting rigid wall, the expressions for the normalized tearing stability index $\hat \Delta' = \Delta'/S^{1/3}$ can be rearranged to calculate the normalized growth rate $\hat \gamma$ in terms of $Q_*=2Q_e$ (for $\tau=1$), $P$, $D$, and $T_i/T_e = \tau$. Note the assumption that $P \sim P_{\phi} \sim P_{\perp}$. The four identified linear regimes are diffusive-resistive (DR), viscous-resistive (VR), semi-collisional (SC), and resistive-inertial (RI). Rearranging the expressions for $\hat \Delta$ derived in Chapter 6 of Ref.~\cite{fitzpatrick_2023} and substituting in this normalized tearing stability index yields corresponding expressions for each linear, normalized growth rate:

\begin{equation}
\begin{aligned}
    \hat\gamma_{DR} = \frac{\Gamma(1/4)}{2 \pi \Gamma(3/4) } \frac{(1+1/\tau)^{1/4} D^{1/2}}{P^{1/4}}\hat \Delta' ,
\label{eq:DR}
\end{aligned}
\end{equation}

\begin{equation}
\begin{aligned}
    \hat \gamma_{VR} = \frac{\Gamma(1/6)}{6^{2/3} \pi \Gamma(5/6) P^{1/6}} \hat \Delta' ,
\label{eq:VR}
\end{aligned}
\end{equation}

\begin{equation}
\begin{aligned}
    \hat \gamma_{SC} = \frac{e^{i \pi /4}}{\pi}\frac{(1 + 1/\tau)^{1/2}D}{Q_*^{1/2} } \hat \Delta' ,
\label{eq:SC}
\end{aligned}
\end{equation}

\begin{equation}
\begin{aligned}
    \hat \gamma_{RI} = \frac{e^{i \pi/2}}{\pi} \frac{(1 + 1/\tau)^{1/2} P^{1/2}}{Q_*} \hat \Delta'.
\label{eq:RI}
\end{aligned}
\end{equation} 

Boundaries for all four linear growth rate regimes can be found in Chapter 6 of Ref.~\cite{fitzpatrick_2023}. Fig.~\ref{fig:Pscan} shows the numerical SLAYER growth rates transition smoothly between three different analytic regimes across a broad range of $P$ values, aligning well with the analytic limits and showing only small offsets in $\hat \gamma$ magnitude. Fig.~\ref{fig:Qscan} shows a similar regime transition with good agreement, but across $Q_*$ instead. 

Scanning $\hat P = P/D^6$ and $\hat Q_* = Q_*/D^4$ in conjunction yields two dimensional linear growth rate maps across all four identified regimes, as can be seen in Figs. \ref{fig:analytic_map} \& \ref{fig:slayer_map}. All 1D and 2D analytic growth rate scans were produced by fixing $\hat \Delta' = \Delta'/S^{1/3} = 0.01$, $D = 2.26$, and $\tau = 1.0$, with the goal of reproducing the map shown in Fig.~6 of Ref.~\cite{fitzpatrick_2022}. All resulting $\hat \gamma$ values were left dimensionless (i.e. not divided by $S^{1/3} \tau_H$). Note that $\mathrm{Re}(\hat \gamma_{RI}) = 0$ implies a purely oscillatory tearing mode in this analytic formalism (for visualization, $
\mathrm{Im}(\hat \gamma_{RI})$ is instead plotted in Fig.~\ref{fig:analytic_map}). Typical tokamak plasmas span the DR (at low field) and VR (at high field) regimes with $\hat Q_* << 1$ \cite{fitzpatrick_2023}. Comparing Figs. \ref{fig:analytic_map} \& \ref{fig:slayer_map}, it is clear that SLAYER agrees well with the analytic results at small $\hat Q_*$. The results diverge more noticeably in the SC regime, with SLAYER showing dependence on both $Q_*$ and $P$, however this discrepancy only occurs at unrealistically high diamagnetic frequencies for a typical tokamak. Additionally, the SLAYER results smoothly connect the disparate analytic regimes, more accurately capturing the physics near regime boundaries.

\begin{figure}[htbp]
\stepcounter{figure}
\refstepcounter{subfigure}\label{fig:Pscan}
\refstepcounter{subfigure}\label{fig:Qscan}
    \centering
    \begin{tikzpicture}
        % --- Subfigure a) ---
        % Set width to 1.0\linewidth to fill the column
        \node[anchor=north west, inner sep=0] (imgA) at (0,0) {%
            \includegraphics[trim={0.7cm 0.5cm 0.0cm 0.0cm}, clip, width=0.5\linewidth]{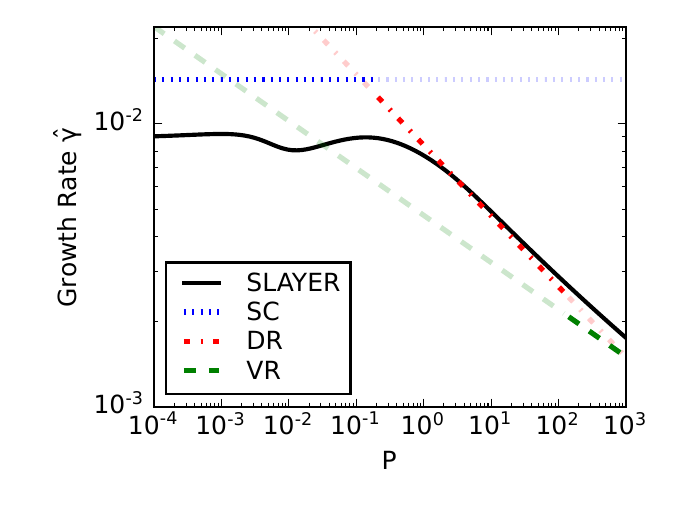}%
        };
        % Label a) anchored to top-left
        % Optional: Add 'text=white' if your plot background is dark
        \node[anchor=north west, font=\bfseries] at (imgA.north west) {a)};

        % --- Subfigure b) ---
        % Anchor the top-left (north west) of this node...
        % ...to the bottom-left (south west) of the previous node (imgA)
        % yshift adds a vertical gap between them
        \node[anchor=north west, inner sep=0, yshift=-0.5cm, xshift=-4mm] (imgB) at (imgA.south west) {%
            \includegraphics[trim={0.5cm 0.5cm 0.0cm 0.0cm}, clip, width=0.5\linewidth]{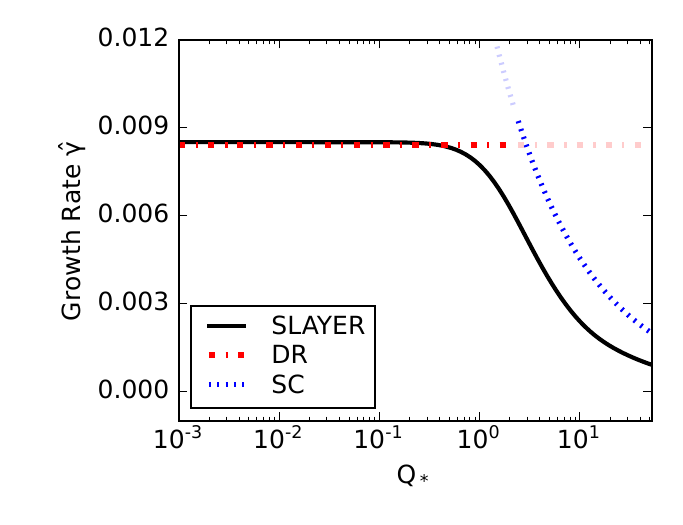}%
        };
        % Label b) anchored to top-left
        \node[anchor=north west, font=\bfseries] at (imgB.north west) {b)};
    \end{tikzpicture}

    \addtocounter{figure}{-1}
    
    \caption{Comparison of SLAYER vs. analytic dimensionless growth rates $\hat{\gamma}$. 
    (a) As viscosity $P$ increases, the linear growth rate regime switches from semi-collisional (SC) to diffusive-resistive (DR) to viscous-resistive (VR). 
    (b) As total normalized diamagnetic rotation $Q_*$ increases, the regime switches from diffusive-resistive (DR) to semi-collisional (SC). 
    Analytic limits are shown as dashed lines. Regions of validity for analytic limits are shown in color, while invalid regions are plotted at low opacity.}
\end{figure}

\begin{figure}[htbp]
    \stepcounter{figure}
    \refstepcounter{subfigure}\label{fig:analytic_map}
    \refstepcounter{subfigure}\label{fig:slayer_map}
    \centering
    \begin{tikzpicture}
        % --- Subfigure a) Analytic Map ---
        \node[anchor=north west, inner sep=0] (imgA) at (0,0) {%
            \includegraphics[width=0.5\linewidth]{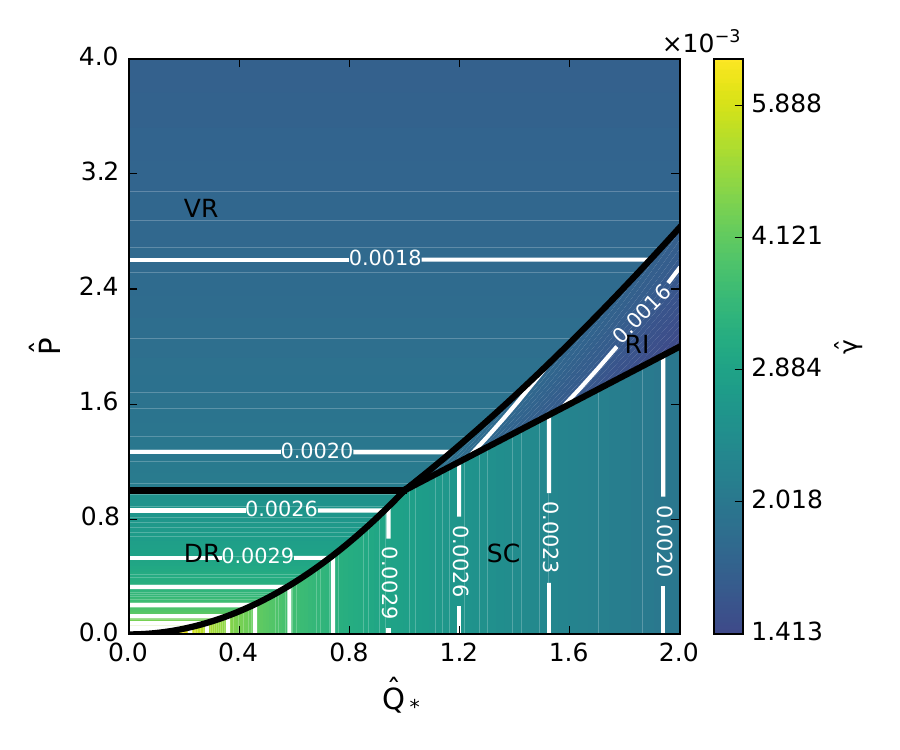}%
        };
        \node[anchor=north west, font=\bfseries] at (imgA.north west) {a)};

        % --- Subfigure b) SLAYER Map ---
        % 'yshift' controls vertical gap
        % 'xshift' is set to 0cm. Change this (e.g., -0.2cm) if the maps don't align perfectly.
        \node[anchor=north west, inner sep=0, yshift=-0.5cm, xshift=0cm] (imgB) at (imgA.south west) {%
            \includegraphics[width=0.5\linewidth]{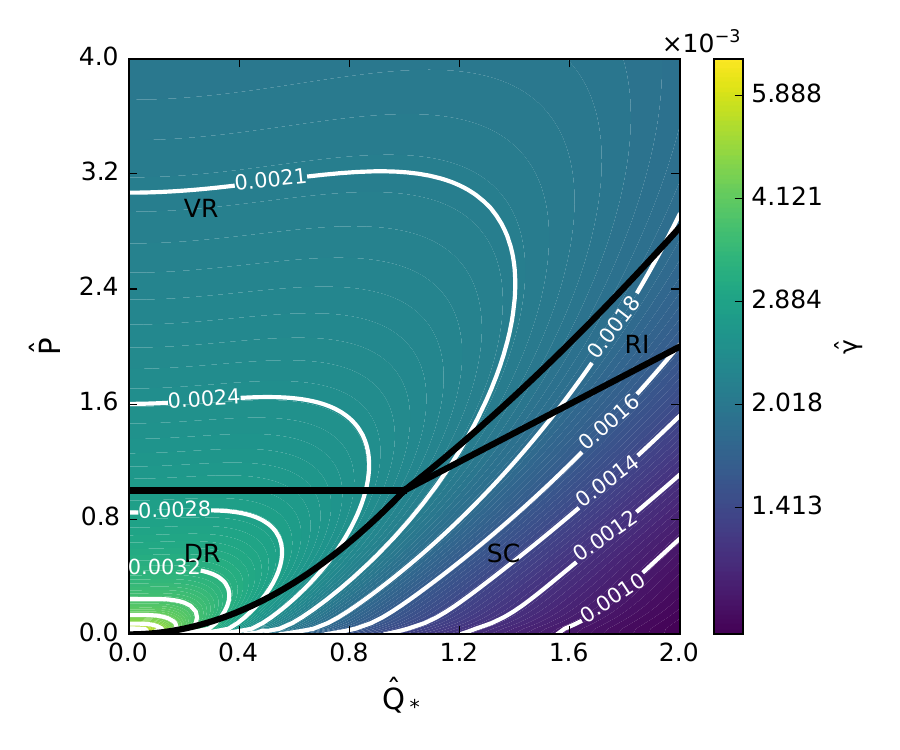}%
        };
        \node[anchor=north west, font=\bfseries] at (imgB.north west) {b)};
    \end{tikzpicture}

    \addtocounter{figure}{-1}
    
    \caption{Full ($\hat Q_*$, $\hat P$) TM growth rate maps across DR, VR, SC, and RI regimes. 
    (a) Analytic map showing discrete linear growth rate regimes as calculated by equations \eqref{eq:DR}--\eqref{eq:RI}. 
    (b) SLAYER map showing smooth transitions between regimes and alignment with analytic theory at tokamak-relevant frequencies (i.e., low $\hat Q_*$).}
    \label{fig:maps}
\end{figure}

\section{\label{sec:TJ_benchmark} Growth Rate Benchmark Against TJ}

Having demonstrated that SLAYER provides numerically reliable growth rates that agree with and smoothly span all analytic regimes, its performance in conjunction with the STRIDE outer region $\Delta'$ can now be investigated on rational surfaces in simple, synthetic tokamak equilibria (hereafter referred to as the STRIDE+SLAYER workflow). Growth rate benchmarks on self-consistent MHD equilibria and kinetic profiles are most easily done using another linear stability code, which is the approach taken in this section.

\subsection{\label{sec:MultiScaling} $\beta$ Scan}

The TJ code calculates linear tearing growth rates of an inverse aspect-ratio ($\epsilon$) expanded tokamak plasma equilibrium using the inner-outer region asymptotic matching technique \cite{fitzpatrick_TJ_2024}. It is therefore an excellent tool for benchmarking this STRIDE+SLAYER workflow. While STRIDE and TJ use differing approaches to calculate the $\Delta'$ matrix, their inner-layer models are equivalent, therefore the resulting growth rates are expected to be roughly identical for a given outer region $\Delta'$.

To benchmark tearing growth rates in a regime where TJ is expected to be valid, a $\beta$ scan was prepared using a circular cross-section equilibrium that replicates the scan shown in Fig.~7 of Ref.~\cite{fitzpatrick_TJ_2025}. This scan covers values of $\beta_N$ from $9.27 \times 10^{-3}$ to $1.56$. TJ was used to output equilibrium and kinetic profile files alongside its tearing growth rates, and these files were then used as inputs to STRIDE and SLAYER for self-consistency. The other equilibrium and kinetic parameters input to TJ are: $q_0 = 1.5$, $q_a = 3.6$, $p_p = 2.0$, $B_0 = 12~\mathrm{T}$, $R_0 = 2~\mathrm{m}$, $a = 0.4~\mathrm{m}$, $n_0 = 3 \times 10^{20}~\mathrm{m}^{-3}$, $T_{\textit{e, edge}} = 100~\mathrm{eV}$, $\alpha$ = 0.5, $Z_{\mathrm{eff}} = 2$, $M = 2$, and $\chi_{\perp} = 0.2~\mathrm{m}^{-3}$ (see Ref.~\cite{fitzpatrick_TJ_2025} for definitions). The perpendicular energy diffusivity $\chi_{\perp}$ is assumed to be spatially uniform. 

%The Newton iteration $\gamma$ root finding procedure used in both codes can fail at very large $\Delta'$ ($> 100$) which occur near ideal stability and $\beta$ limits, therefore scan points where the root finding failed have been removed.

\begin{figure}[h]
\centering\includegraphics[trim={0.5cm 0.5cm 0.0cm 0.0cm},clip,width=0.5\linewidth]{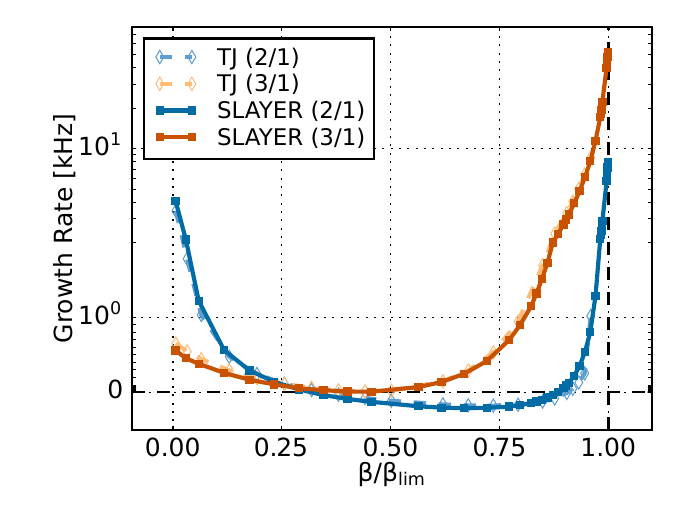}\caption{STRIDE+SLAYER vs TJ $\beta$ scan benchmark, showing good alignment in the 2/1 and 3/1 growth rates calculated by both codes. \label{fig:betascan_gammas}}
\end{figure}

In Fig.~\ref{fig:betascan_gammas}, good alignment is seen between the TJ and STRIDE+SLAYER 2/1 and 3/1 growth rates.  STRIDE and TJ see different $\beta$ limits (similar to previously investigated discrepancies between PEST-3 \cite{Pletzer_1994_JCP} and RDCON \cite{Wang_RDCON_2020}), therefore the scans are plotted across the fractional $\beta$ limit seen by each code. Scan points near the $\beta$ limit ($\Delta' > 100$) where the TJ failed to solve for the growth rate have been removed.

\begin{figure}[h]
\centering\includegraphics[trim={0.5cm 0.5cm 0.0cm 0.0cm},clip,width=0.5\linewidth]{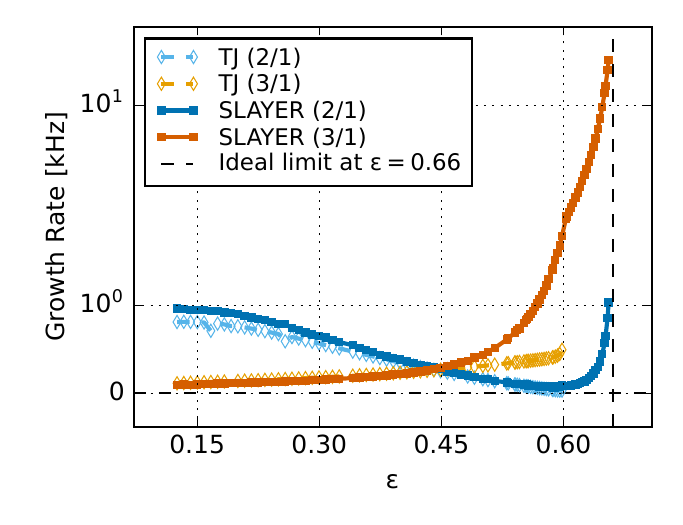}\caption{Inverse aspect-ratio scan showing STRIDE+SLAYER growth rate divergence from TJ at high $\epsilon$. \label{fig:epsscan_gammas}}
\end{figure}

\begin{figure}[h]
\centering\includegraphics[trim={0.5cm 0.5cm 0.0cm 0.0cm},clip,width=0.5\linewidth]{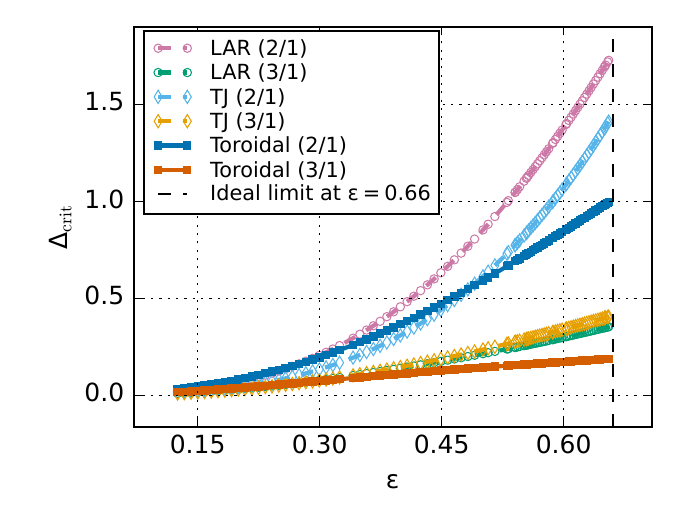}\caption{Toroidal $\Delta_\mathrm{crit}$ used in SLAYER (``Toroidal''), large aspect-ratio approximated $\Delta_\mathrm{crit}$ used in TJ (``TJ''), and original Ref.~\cite{curvature_Lutjens_2001} $\Delta_\mathrm{crit}$ (``LAR'') compared across $\epsilon$ scan. \label{fig:epsscan_dcs}}
\end{figure}

\subsection{\label{sec:MultiScaling} $\epsilon$ Scan}

Because TJ uses an $\epsilon$-expansion approach that may not capture higher order geometric effects of shaping in the outer region to the same fidelity as STRIDE, STRIDE and TJ may be expected to differ at large $\epsilon$ where the $\epsilon$-expansion is less valid or insufficiently high order. For this $\epsilon$ scan, all input equilibrium and kinetic parameters were held fixed as detailed in the $\beta$ scan, except $\beta_N$ which was set to $9.27 \times 10^{-3}$ (the minimum value from that scan), and $a$ to $1.0$ m. $R_0$ was then scanned from 8 m to 1.52 m, corresponding to $\epsilon$ values of 0.125 to 0.658. As seen in Fig.~\ref{fig:epsscan_gammas}, at low $\epsilon$, both codes follow the same trend in growth rates with only a small offset in $\Delta'$ values that propagates into $\gamma$. As $\epsilon$ increases past 0.45, the STRIDE $\Delta'$ increases as it approaches an ideal limit, driving up both the SLAYER 2/1 and 3/1 growth rates. However, the TJ $\Delta'$ (and thus $\gamma$) values diverge noticeably in slope and trend for $\epsilon > 0.45$ before reaching a different ideal limit at $\epsilon = 0.60$, suggesting that STRIDE may capture trends in $\Delta'$ unseen by TJ in low aspect-ratio (high $\epsilon$) tokamak plasmas. Additionally, discrepancies would likely appear earlier in $\epsilon$ in more poloidally shaped plasmas. 

In Fig.~\ref{fig:epsscan_dcs}, the TJ large aspect-ratio approximated $\Delta_\mathrm{crit}$ values are compared against the toroidal $\Delta_\mathrm{crit}$ used by SLAYER. Both are in reasonable agreement, though there is some offset at typical tokamak $\epsilon$ values ($\sim0.3$ to $0.6$). For comparison, the original large aspect-ratio approximated $\Delta_\mathrm{crit}$ that included thermal conduction effects and was derived by Ref.~\cite{curvature_Lutjens_2001} is compared against TJ and SLAYER, denoted ``LAR'' in this figure. The toroidal $\Delta_\mathrm{crit}$ predicts less stabilization than both large aspect-ratio approximated values across most $\epsilon$ values in this scan.

%This Lutjens et al. $\Delta_\mathrm{crit}$ is much larger and possibly an over-prediction when considered against the more recent derivations of $\Delta_\mathrm{crit}$ used in SLAYER and TJ. 

\section{\label{sec:D3D_example} Growth Rate Evolution with Changing Toroidal Current Profile in a Shaped Plasma}

In order to investigate the tearing stability of experimental tokamak discharges using this STRIDE+SLAYER workflow, confidence is needed in the workflow's ability to handle growth rate solves in a more realistic parameter space (i.e. $T_e \neq T_i$, in shaped equilibria, and with reverse-shear regions in the toroidal current profile $j_{\phi}$). To investigate this, the ``TokaMaker'' Grad-Shafranov solver was used, which is part of OpenFUSIONToolkit \cite{tokamaker_Hansen_2024}. The chosen baseline synthetic equilibrium is representative of typical H-modes in DIII-D \cite{Carlos_RMP_2024}, with the following set of parameters: $I_p = 1.4$ MA, $B = 2$ T, elongation $\kappa = 1.7$, triangularity $\delta = 0.4$, $q_0 = 1.1$, $q_{95} = 3.7$, and $\beta_N = 1.2$ $\%$. Parameterized\footnote{See function \texttt{Hmode$\_$profiles()} in \url{https://omfit.io/_modules/omfit_classes/utils_fusion.html} for profile parameterization.} H-mode-like kinetic profiles are used with $n_{e,core} = 5.5 \times 10^{19}~\mathrm{m}^{-3}$, $n_{e,ped} = 3.3 \times 10^{19}~\mathrm{m}^{-3}$, $T_{e,core} = 2.5$ keV, $T_{e,ped} = 1.0$ keV, $n_{i,core} = 4.5 \times 10^{19}~\mathrm{m}^{-3}$, $n_{i,ped} = 2.7 \times 10^{19}~\mathrm{m}^{-3}$, $T_{i,core} = 5.0$ keV, and $T_{i,ped} = 1.5$ keV.

\begin{figure}[h]
\centering\includegraphics[width=0.5\linewidth]{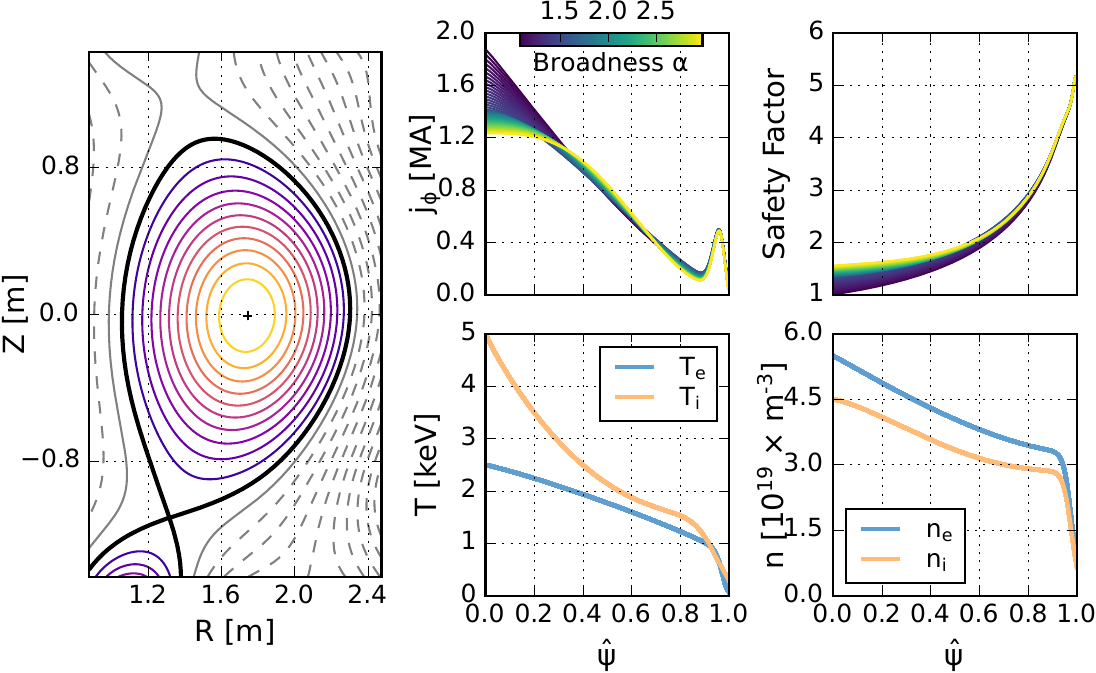}\caption{TokaMaker synthetic equilibrium, fixed density and temperature profiles, and scanned $j_\phi$ and safety factor profiles used to increase the gradient of $j_\phi$ and therefore $\Delta'$ on the 2/1 and 3/1 surfaces. \label{fig:jscan_profs}}
\end{figure}

\begin{figure}[h]
\centering\includegraphics[trim={0.5cm 0.5cm 0.0cm 0.0cm},clip,width=0.5\linewidth]{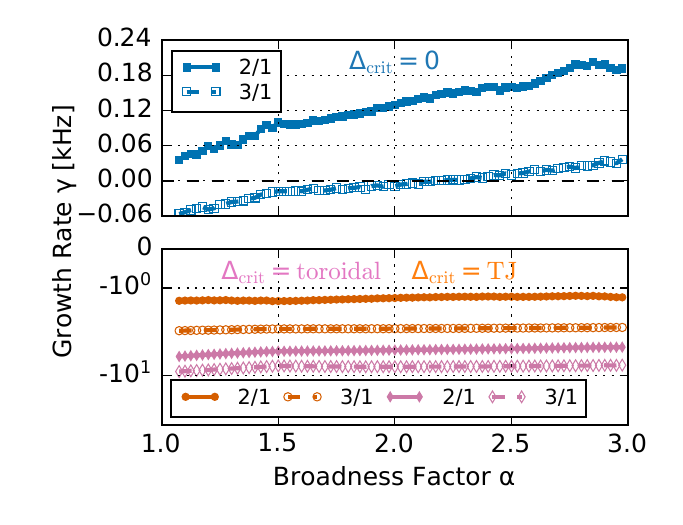}\caption{2/1 and 3/1 growth rates across the toroidal current profile scan shown in Fig.~\ref{fig:jscan_profs}. The square marker cases use $\Delta_\mathrm{crit}=0$, the circle marker cases use the TJ $\Delta_\mathrm{crit}$, and the diamond marker cases use the toroidal $\Delta_\mathrm{crit}$.  \label{fig:jscan_gammas}}
\end{figure}

The parallel bootstrap current ($j_{\mathrm{BS}}$) profile is calculated using the Sauter formula \cite{Sauter_bootstrap_1999}, and the inductive $j_\phi$ profile is parameterized as $j_\phi({\psi_N}) = \left(1 - {\psi_N}^{\alpha}\right)^{\gamma}$. Both $j_\phi$ and $j_{\mathrm{BS}}$ are summed, and the core $j_\phi$ broadness is increased using the inner exponent $\alpha$ (hereafter referred to as a ``broadness factor'') as can be seen in Fig.~\ref{fig:jscan_profs}. %As shown in Fig.~\ref{fig:jscan_gammas}, increasing core $j_\phi$ broadness drives the (3/1) $\Delta'$ from stabilizing to destabilizing, which proportionally drives the (3/1) growth rate unstable. 
Growth rates with and without the inclusion of Glasser stabilization are both presented in Fig.~\ref{fig:jscan_gammas}. Excluding stabilization effects yields unstable 2/1 growth rates across the broadness scan, whereas the 3/1 surface is driven from stability to instability as the gradient of $j_\phi$ and therefore $\Delta'$ both increase. For these DIII-D-like profiles parameters, the calculated values of $\Delta_\mathrm{crit}$ using both the TJ and toroidal definitions are overwhelmingly stabilizing, and without analysis of experimental data it remains unclear which form of the dispersion relation better aligns with observed classical TM onset. These initial studies using analytic, shaped equilibria with realistic profiles prove the STRIDE+SLAYER workflow to be a robust and interpretable tool that is now ready for application to and validation on experimental time-series kinetic equilibria from present day tokamaks.

\section{\label{sec:summary} Summary and Future Work}

The basic capabilities of a new classical TM stability calculation workflow have been demonstrated, combining successful analytic and code-to-code benchmarks before application to more realistic synthetic H-mode-like equilibria. This workflow matches high physics fidelity inner region solutions for $\Delta$ (calculated via SLAYER) with the ideal MHD toroidal $\Delta'$ (calculated via STRIDE) on all uncoupled rational surfaces in a given equilibrium. This approach does not posses the fidelity of ``straight-through'' resistive MHD codes such as M3D-C1, and additionally neglects pressure gradients and parallel flow inside the slab. However, these latter approximations are justifiable in present-day and future devices with low to moderate $\beta$ and $\sim$keV temperatures. Using these approximations in the original four field slab model yields an ODE for the slab's inner region $\Delta$ that is well behaved and rapidly solved, allowing for quick and robust TM dispersion relation root finding for the growth rate $\gamma$ and mode rotation frequency $\omega$. The asymptotically matched TM growth rate formalism was further extended in this work by including a toroidal proxy for Glasser stabilization on each rational surface, which yields an effective tearing stability index $\Delta_\mathrm{eff} = \Delta' - \Delta_\mathrm{crit}$ used for all inner-outer region matching. 

In all cases analyzed so far, this toroidal $\Delta_\mathrm{crit}$ is within a factor of $\sim$$2$ from the LAR approximated $\Delta_\mathrm{crit}$  used in the TJ code. However, the values obtained from both definitions appear overwhelmingly dominant in equilibria more relevant to modern tokamaks, such as those in Figs. \ref{fig:jscan_profs} \& \ref{fig:jscan_gammas}. The inclusion of $\Delta_\mathrm{crit}$ in the dispersion relation and its behavior in experimental equilibria requires further analysis, since the sign of $\Delta_\mathrm{eff}$ defines the stability of the classical tearing mode, and this stability cannot be altered in typical core plasma conditions by any other inner region physics presently contained within SLAYER. 

Additionally, the inclusion of higher order effects such as surface-to-surface coupling  \cite{Brennan_coupling_2006} and the effect of a resistive wall \cite{Finn_1995_PoP} may prove necessary for further aligning linear TM theory with experimental results, and their implementation and analysis would benefit the disruption physics community. The effects of two-surface, three-surface, etc. coupling on TM growth or stabilization remain largely unexplored, and it is additionally unclear whether coupling, this work's adopted proxy for Glasser stabilization, or their combined effects most realistically align with experimentally observed TM onset.

In conclusion, this STRIDE+SLAYER TM stability workflow offers a robust, toroidally self-consistent, and open source analysis tool \cite{doecode_23325} that can be rapidly deployed on both analytic equilibria and experimental results. While stochastic NTMs will likely remain a threat in the discharge flat-top of present-day and future devices, the STRIDE+SLAYER classical TM workflow can aid in the investigation and design of safe discharge trajectories of dynamic phases such as ramp-up and ramp-down, as well as mapping parameter regions of greatest classical TM stability for full-power operation. 

\section{\label{sec:Acknowledgments}Acknowledgments}

The authors thank Dylan Brennan and Richard Fitzpatrick for helpful discussions on linear tearing mode theory, as well as support from Prof. Fitzpatrick when using TJ for code-to-code benchmarking. This work was supported by the U.S. Department of Energy Office of Science Office of Fusion Energy Sciences under Award DE-SC0022272.

\textbf{Disclaimer:} This report was prepared as an account of work sponsored by an agency of the United States Government. Neither the United States Government nor any agency thereof, nor any of their employees, makes any warranty, express or implied, or assumes any legal liability or responsibility for the accuracy, completeness, or usefulness of any information, apparatus, product, or process disclosed, or represents that its use would not infringe privately owned rights. Reference herein to any specific commercial product, process, or service by trade name, trademark, manufacturer, or otherwise does not necessarily constitute or imply its endorsement, recommendation, or favoring by the United States Government or any agency thereof. The views and opinions of authors expressed herein do not necessarily state or reflect those of the United States Government or any agency thereof.

\section*{Conflict of Interest}
The authors have no conflicts to disclose.

\section*{Data Availability}
The TJ equilibrium files used as inputs for the STRIDE+SLAYER workflow and output SLAYER growth rate data corresponding to Figs. \ref{fig:betascan_gammas}, \ref{fig:epsscan_gammas}, and \ref{fig:epsscan_dcs} are openly available in Zenodo at \url{https://doi.org/10.5281/zenodo.18329503}.

\appendix
\section{\label{appen:plasma_params}Additional SLAYER Plasma Parameters}

See Table~\ref{tab:more_normalized_variables}.

% Pucella, Brennan, Alan Turnbull
% adturnbull53@gmail.com

\begin{table}[ht]
\centering
\caption{Additional parameters required for calculation of SLAYER quantities in Table~\ref{tab:normalized_variables}. Electron and ion temperatures $T_e$ and $T_i$ are in units of eV. $q$ is the elementary charge. $\mu$ is the main ion to proton mass ratio, while $\mu_i$ is the phenomenological ion viscosity.}
\renewcommand{\arraystretch}{1.75} % 1.5 gives 50% more height
\begin{tabular}{l|l|p{8cm}}
\multicolumn{1}{c}{\textbf{Variable}} & \multicolumn{1}{c}{\textbf{Definition}} & \multicolumn{1}{c}{\textbf{Description}} \\
\hline\hline
$\ln\Lambda$ & $24 + 3.0\ln(10) - 0.5\ln(n_e) + \ln(T_e)$ & Coulomb logarithm\\
\hline
$\tau$ & $T_i/T_e$ & Ion to electron temperature ratio\\
\hline
$\eta$ & $1.65 \times 10^{-9} \ln\Lambda/(T_e/10^3)^{1.5}$ & Spitzer resistivity \\
\hline
$\rho$ & $\mu m_p n_e$ & Mass density\\
\hline
$\tau_{ee}$ & $\frac{6\sqrt{2}\pi^{1.5} \varepsilon_0^2 m_e^{0.5} T_e^{1.5}}{\ln\Lambda \cdot q^{2.5} n_e}$ & Electron-electron collision time\\
\hline
$\sigma_{\parallel}$ & $\frac{\sqrt{2} + 13(Z_{\mathrm{eff}}/4)}{Z_{\mathrm{eff}}(\sqrt{2} + Z_{\mathrm{eff}})} \cdot \frac{n_e q^2 \tau_{ee}}{m_e}$ & Parallel conductivity\\
\hline
$\rho_s$ & $1.02 \times 10^{-4} \sqrt{\mu T_e}/B_t$ & Ion Larmor radius \\
\hline
$d_i$ & $\sqrt{\mu m_p/(n_e e^2 \mu_0)}$ & Collisionless ion skin depth\\
\hline
$\tau_H$ & $R_0 \sqrt{\mu_0 \rho}/(n_n s B_t)$ & Alfvén time across rational surface\\
\hline
$\tau_R$ & $\mu_0 r_s^2 \sigma_{\parallel}$ & Resistive timescale \\
\hline
$\tau_V$ & $r_s^2 \rho/\mu_i$ & Viscous timescale (phenomenological) \\
\hline
$\beta$ & $(5/3)\mu_0 n_e q (T_e+T_i)/B_t^2$ &  Plasma $\beta$ \\
\hline
$c_\beta$ & $\sqrt{\beta/(1 + \beta)}$ & Dimensionless pressure parameter \\
\hline
$d_\beta$ & $c_\beta d_i$ & Ion sound radius \\
\hline
$S$ & $\tau_R/\tau_H$ & Lundquist number \\
\hline
$K$ & $\kappa/\eta$ & $\kappa=$ thermal conductivity \\
\hline
\end{tabular}
\renewcommand{\arraystretch}{1}

\label{tab:more_normalized_variables}
\end{table}

\printbibliography

%\end{multicols}

\end{document}